\documentclass[%
amsmath,amssymb,
aps,
mph,%
reprint,%
tightenlines,
notitlepage,
longbibliography,
superscriptaddress,
floatfix,
nofootinbib,
]{revtex4-2}

\usepackage{nccmath}
\usepackage{setspace}
\usepackage{booktabs}
\usepackage{graphicx}
\usepackage[table,xcdraw,pdftex,dvipsnames]{xcolor}
\usepackage{dcolumn}
\usepackage{bm}
\usepackage{siunitx}
\usepackage{subcaption}
\usepackage{amsmath}
\usepackage{amssymb}
\usepackage{amsthm}
\usepackage{booktabs}
\usepackage{array}
\usepackage{multirow} 
\usepackage{nicematrix} 
\usepackage{ragged2e}
\usepackage{textcomp}
\usepackage{enumitem}
\usepackage{mathtools}
\usepackage{hyperref}
\usepackage[toc,page]{appendix}
\usepackage{lipsum}
\usepackage{cleveref} 
\usepackage[T1]{fontenc} 
\usepackage{comment}

\usepackage{tikz}
\usetikzlibrary{fit,shapes.misc}

\DeclareCaptionStyle{mystyle}
  {format=plain,%
    justification=raggedright,
    singlelinecheck=false, 
  }
\newcommand{\myrowsep}{1.25} 
\newcommand{\mytablefontsize}{\footnotesize} 

\newcolumntype{L}{>{$}l<{$}}
\newcolumntype{C}{>{$}c<{$}}
\newcolumntype{R}{>{$}r<{$}}

\usepackage[mathlines]{lineno}
\modulolinenumbers[5]

\usepackage{xargs}
\usepackage[colorinlistoftodos,prependcaption]{todonotes}
\newcommandx{\unsure}[2][1=]{\todo[linecolor=red,backgroundcolor=red!25,bordercolor=red,#1]{#2}}
\newcommandx{\change}[2][1=]{\todo[linecolor=blue,backgroundcolor=blue!25,bordercolor=blue,#1]{#2}}
\newcommandx{\info}[2][1=]{\todo[linecolor=OliveGreen,backgroundcolor=OliveGreen!25,bordercolor=OliveGreen,#1]{#2}}
\newcommandx{\improvement}[2][1=]{\todo[linecolor=Plum,backgroundcolor=Plum!25,bordercolor=Plum,#1]{#2}}
\newcommandx{\thiswillnotshow}[2][1=]{\todo[disable,#1]{#2}}

\usepackage{etoolbox}
\patchcmd{\section}
  {\centering}
  {\raggedright}
  {}
  {}
\patchcmd{\subsection}
  {\centering}
  {\raggedright}
  {}
  {}

\newcommand{\EXP}[2]{\textrm{EXP}({#1};{#2})}
\newcommand{\GAUSS}[2]{\mathcal{N}({#1};{#2})}
\newcommand{\EQ}[1]{Equation~{#1}}
\newcommand{\FIG}[1]{Figure~{#1}}
\newcommand{\SECT}[1]{Section~{#1}}

\begin{document}

\title{An analytic, moment-based method to estimate orthopositronium lifetimes in positron annihilation lifetime spectroscopy measurements}

\author{Lucas Berens}
\email{Author to whom correspondence should be addressed. Electronic mail: lberens@uchicago.edu}

\affiliation{Department of Radiology, The University of Chicago, Chicago, IL, USA}
\affiliation{Department of Radiation and Cellular Oncology, The University of Chicago, Chicago, IL, USA}

\author{Isaac Hsu}
\affiliation{Department of Computer Science, University of Minnesota, Minneapolis, MN, USA}

\author{Chin-Tu Chen}
\affiliation{Department of Radiology, The University of Chicago, Chicago, IL, USA}

\author{Howard Halpern}
\affiliation{Department of Radiation and Cellular Oncology, The University of Chicago, Chicago, IL, USA}

\author{Chien-Min Kao}
\affiliation{Department of Radiology, The University of Chicago, Chicago, IL, USA}


\begin{abstract}
\noindent The presence of tumor hypoxia is known to correlate with poor patient prognosis. Measurement of tissue oxygen concentration can be challenging, but recent advancements using positron annihilation lifetime spectroscopy (PALS) in three-dimensional positron emission tomography (PET) scans have shown promise for hypoxia detection. In this work, a novel method for estimating the orthopositronium lifetime in PALS is presented. This method is analytical and uses moments of the time-difference histogram from photon arrival times. For sufficient statistical power, the method produces monotonic, stable estimates. For cases with a lower number of photon counts, the method was characterized and solutions are presented to correct for bias and estimation variability. 
\end{abstract}
\maketitle

\section{Introduction}
\label{sec:Intro}
Positronium lifetime imaging (PLI) is a novel augmentation
for positron emission tomography (PET) that may allow PET scans to extract tissue oxygenation information, including hypoxic locations, in addition to the specific biochemical properties the employed PET tracer is responsive to. Recently, Moskal et al. has identified the positronium lifetime as a possible metric for hypoxia \cite{moskal2022positronium_review}. The same group has also designed and built a novel PET scanner for generating positronium lifetime images \cite{moskal2021positronium}. Other recent work includes that of Shibuya et al. in which the correlation of oxygen concentration with positronium lifetimes was rigorously established \cite{shibuya2020oxygen}. 

PLI makes use of measurements of the lifetime of positronium (Ps), which are short-lived bound states between a positron emitted from nuclear decay and an electron from the environment. In clinical PET circumstances, approximately 40\% of the released positrons lead to this electron-positron bound state \cite{shibuya2020oxygen}. The other 60\% annihilate with electrons without forming positronium. These produce direct annihilations (DA), and predominantly produce two coincident \SI{511}{keV} photons with a lifetime of about \SI{388}{ps}~\cite{moskal2021positronium}.

There are two types of positronium, orthopositronium (o-Ps) where the electron and positron spins are parallel,  and parapositronium (p-Ps) where the spins are anti-parallel. Both are formed within \SI{5}{ps} of the release of the positron after it has thermalized with the environment. In tissue, approximately 75\% of positronium is o-Ps and 25\% is p-Ps. Both o-Ps and p-Ps are inherently unstable and eventually their constituent electron and positron annihilate with each other without interacting with their environment. In such self annihilations, p-Ps produces two \SI{511}{keV} photons and has a short lifetime of approximately \SI{125}{ps}. On the other hand, o-Ps produces three coincidence photons\footnote{In theory, p-Ps self-annihilation is permitted to produce $n=2, 4, 6,\cdots$ coincidence photons but $n=2$ predominates. Likewise, o-Ps self-annihilation is permitted to produce $n=3,5,7,\cdots$ coincidence photons but $n=3$ predominates. In direct annihilations, $n\geq2$ photons can be created since any number of photons greater than $n=2$ can allow for momentum conservation.} and has a long lifetime of approximately \SI{142} {ns}~\cite{moskal2021positronium_review}.

As a result, o-Ps has the opportunity to interact with electrons that are present in its surroundings and to decay before self-annihilation, thereby shortening the lifetime to  1-\SI{15}{ns} \cite{moskal2021positronium_review}. These interactions include a spin-exchange with an unpaired electron that converts o-Ps to p-Ps where it quickly decays to produce two \SI{511}{keV} photons, and a pick-off event where the positron annihilates with a local electron (also most likely to produce two \SI{511}{keV} photons). The most dominant of these interactions is the spin-exchange. In tissue, the o-Ps lifetime is between 1-\SI{3}{ns} and importantly its value depends on local properties that affect the strength of the spin-exchange interaction \cite{shibuya2020oxygen}. For example, molecular oxygen is paramagnetic and can readily cause o-Ps decay, since o-Ps readily decays by interacting with unpaired electrons.

If we therefore choose an isotope that possesses a nontrivial decay mode in which a prompt gamma and a positron are released at essentially the same time, we may measure the arrival time difference between the prompt gamma and coincident \SI{511}{keV} photons. A metric for local tissue oxygenation may then be derived. Specifically, the histogram $S(\Delta t)$ of  these differences in arrival times $\Delta t$ can be modeled as the sum of multiple convolutions of exponential distributions and Gaussian distributions \cite{moskal2021positronium}. We consider here lifetime components from o-Ps, p-Ps, and DA as they have sufficiently distinct lifetimes and intensities to be modeled separately. Each is represented by one exponential distribution that models the probability distribution for the decay, convolved with a Gaussian distribution that models the statistical uncertainty in time measurement. Therefore, the full model for $S(\Delta t)$ can be written as
\begin{equation}\label{eq:signal}
        S(\Delta t) = b + \sum^{3}_{i=1} I_{i}\
        \EXP{\Delta t}{\lambda_i} \ast
        \GAUSS{\Delta t-t_0}{\sigma_i},
\end{equation}
where $\ast$ denotes convolution, $I_i$, $\lambda_i,$ and $\sigma_i$ are  the intensity, decay-rate constant, and standard deviation of the time-measurement uncertainty associated with component $i$ for $i=1,2,3$, $b\geq 0$ accounts for the presence of background events, EXP is defined as 
\begin{equation}
    \EXP{t}{\lambda}=\lambda\exp\{-\lambda t\}\cdot u(t),
\end{equation}
in which $u(t)$ is the unit step function defined by $u(t)=1$ for $t\geq 0$ and $u(t)=0$ for $t<0$, and
\begin{equation}
    \GAUSS{t}{\sigma} = \left(\sqrt{2\pi}\sigma\right)^{-1}\, \exp\{-t^2/2\sigma^2\}.
\end{equation}
The lifetime $\tau_i$ of component $i$ is related to $\lambda_i$ by $\tau_i=1/\lambda_i$. In Equation {\ref{eq:signal}}, $t_0$ is introduced to allow for an offset in time measurement. We wish to estimate $\tau$ associated with o-Ps from a measurement of $S(\Delta t)$ for providing a metric for certain tissue property. Typically, it is reasonable to take $\sigma_i=\sigma,\,\forall i$ and $t_0=0$.
Without loss of generality, we can take the components $i=1,2,3$ to be those from
the p-Ps decay, DA, and o-Ps decay, respectively, and hence $\lambda_1>\lambda_2>\lambda _3.$

While PLI is emerging for the purpose of producing full three-dimensional images of the o-Ps lifetime, this work is concerned with only the single-dimensional case represented by Equation \ref{eq:signal}. For distinction, we therefore refer to this method as one for \textit{positron annihilation lifetime spectroscopy} (PALS) measurements. In the discussion we will comment on using this method for imaging in general, and will consider it in-depth in a separate paper. 

Presented here is a novel analytic method for estimating the o-Ps lifetime from $S(\Delta t)$. Current methods for this task include fitting a single exponential distribution to the time-difference histogram \cite{shibuya2020oxygen}, and fitting a reduced full model given by \EQ{\ref{eq:signal}} by assuming known values for certain parameters to the histogram \cite{moskal2021positronium}. Additionally, Shibuya et al. has proposed an inverse Laplace transform method to distinguish between positronium lifetimes while merging voxels for better statistics \cite{shibuya2022using}. Their method is able to discern similar lifetime values, but still employs curve-fitting.  In comparison with these curve-fitting methods, the new analytic method in this work is more computationally efficient, which is an important consideration for future application of the method for PLI in which lifetimes need to be obtained for a large number of voxels in a clinical setting. In addition, under mild and realistic conditions the analytic method is not sensitive to the unknown lifetimes for DA and p-Ps, nor to the time-measurement uncertainty $\sigma$. Generally, this is the not case with curve-fitting methods.

\section{Materials and Methods}
\label{sec:methods}

    \subsection{Derivation of the method and justifications}
    \label{subsec:derivation}
    
    The $n$th moment of a function $f(t)$ is defined by
    \begin{equation}\label{eq:moment-defn}
        \mu_n\{f(t)\} = \int_{-\infty}^{\infty} t^n f(t) dt
    \end{equation}
    if the integral exists.
    The proposed analytic method is based on the following observations.

    \begin{enumerate}
        \item The $n$th moment of $\EXP{t}{\lambda}\}$ exists for all $\lambda>0$ and is given by
            \begin{equation}\label{eq:moment-EXP}
            \mu_n\{\EXP{t}{\lambda}\} = \frac{n!}{\lambda^n}.
            \end{equation}
        \item Let $g(t;\lambda,\sigma) = \EXP{t}{\lambda}\ast \GAUSS{t}{\sigma}.$
        For sufficiently large $\lambda$ (with respect to $\sigma$) and $n$,
        it can be shown that, for any $s>0$,
            \begin{equation}\label{eq:moment-EMG}
            \mu_n\{e^{-st} g(t;\lambda,\sigma)\} \approx c\cdot\mu_n\{\EXP{t}{s+\lambda}\},
            \end{equation}
        where $c=\lambda/(s+\lambda) e^{\sigma^2\lambda^2/2}$ is independent of $n$.
        \item Applying the above to \EQ{\ref{eq:signal}} with $\sigma_i=\sigma$ and $t_0=0$, we get
        \begin{equation}\label{eq:moment-att-PAL}
            \mu_n\{e^{-s\Delta t}\tilde{S}(\Delta t)\} \approx c \cdot n! \cdot \sum_{i=1}^{3}
            \frac{I_i}{(s+\lambda_i)^n},
        \end{equation}
        where $\tilde{S}(\Delta t)=(S(\Delta t)-b)\cdot u(\Delta t)$.
        \item  Since $\lambda_3$ is assumed to be smaller than $\lambda_1$
        and $\lambda_2$, the $i=3$ term dominates the sum
        in \EQ{\ref{eq:moment-att-PAL}} for large values of $n$, yielding
        \begin{equation}\label{eq:moment-att-PAL1}
            R(n+1,s) \equiv \frac{\mu_{n+1}\{e^{-st}\tilde{S}(\Delta t)\}}{\mu_{n}\{e^{-st}\tilde{S}(\Delta t)\}} \approx
            \frac{n+1}{s+\lambda_3}.
        \end{equation}
        Therefore, with an adequately large $n$ we have
        \begin{equation}\label{eq:estimate_lambda3}
            \tau_3 \approx \frac{R(n+1,s)}{n+1-sR(n+1,s) + \delta},
        \end{equation}
        where the small positive value $\delta$ has been added to the denominator to control observed estimation variability for small values of $\tau_3$. 
    \end{enumerate}
    
\begin{figure}
        \centering
        \includegraphics[width=\linewidth]{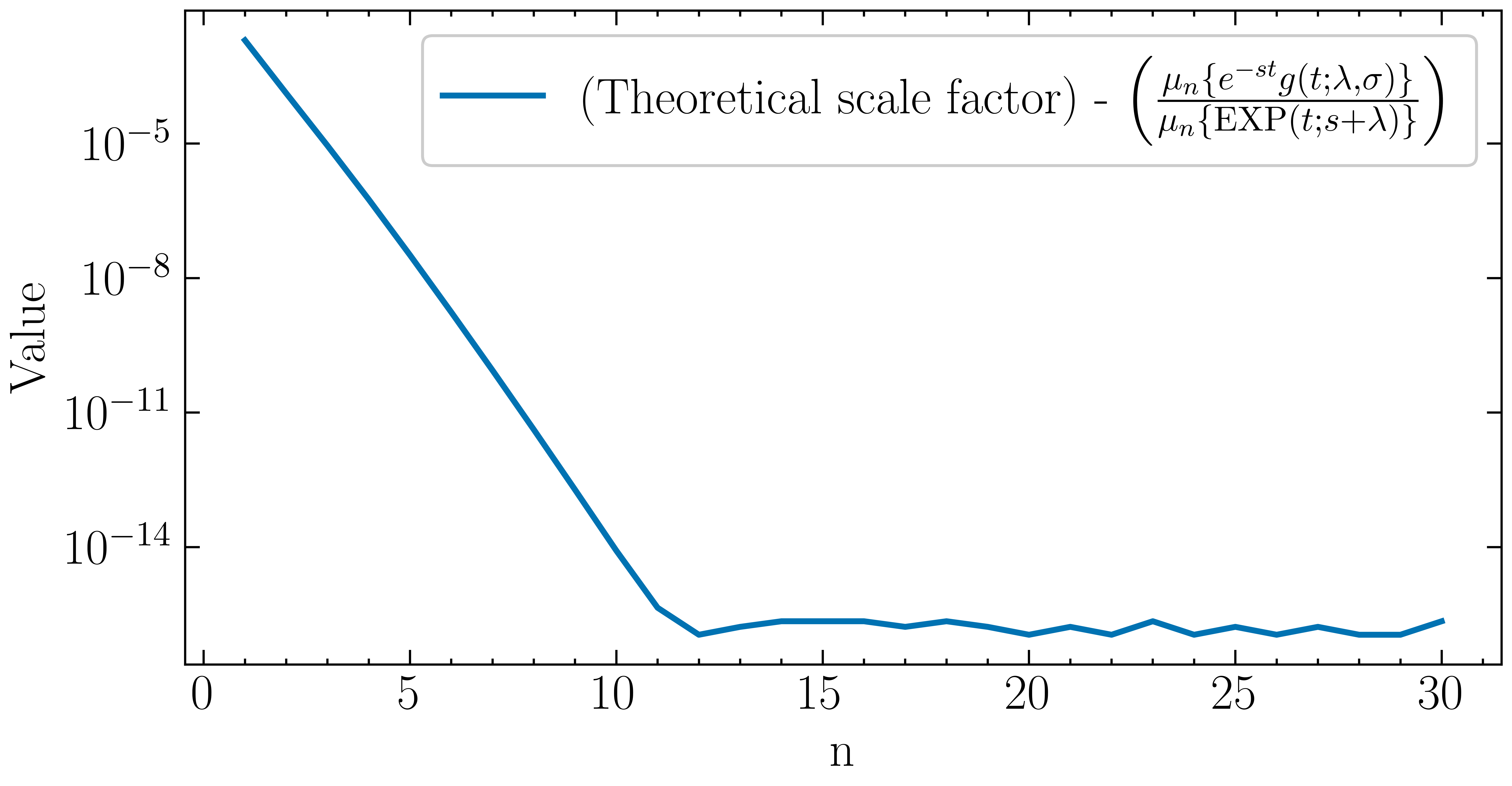}
        \caption{The scale factor $c$ in Equation \ref{eq:moment-EMG}, from which the ratio ${\mu_n\{e^{-st} g(t;\lambda,\sigma)\}}/{\mu_n\{\EXP{t}{s+\lambda}\}}$ has been subtracted. Their difference has a maximum of \SI{2.01e-03}{}, which decreases as $n$ increases until $n=12$ where the mean difference becomes \SI{1.6e-16}{}. }
        \label{ratioEMG_EXP}
    \end{figure}

    The plot in Figure \ref{ratioEMG_EXP} demonstrates \EQ{\ref{eq:moment-EMG}} numerically for $s=1$ and several selected values for $n$ and $\lambda$. Derivations of \EQ{\ref{eq:moment-EXP}} and \ref{eq:moment-EMG} are given in the Appendix.

    For the constant values we used in Equation \ref{eq:signal}, we reference Moskal et al. \cite{moskal2021positronium}. These values are  $\tau_1=1/\lambda_1\approx$ \SI{0.125}{ns} and $I_1\approx 0.1$ for p-Ps, $\tau_2=1/\lambda_2\approx$ \SI{0.388}{ns} and $I_2\approx 0.6$ for DA, and $\tau_3=1/\lambda_3\approx$ 1-\SI{10}{ns} and $I_3\approx 0.3$ for o-Ps. The coincidence resolving time (CRT) was chosen to be \SI{600}{ps} full width at half maximum (FWHM) from  current state-of-the-art time-of-flight (TOF) PET systems \cite{conti2011improving, moskal2021positronium}, yielding $\sigma\lesssim$\SI{220}{ps}. Therefore, for PALS and PLI measurements, the assumptions leading to \EQ{\ref{eq:estimate_lambda3}} can be verified.\footnote{Given the CRT in FWHM, if time measurements made by all channels are independent, the standard deviation of the single-channel time measurement of a TOF PET system is $\sigma_1=\sigma_\mathrm{TOF}/\sqrt{2}$ where $\sigma_\mathrm{TOF}=\mathrm{CRT}/2.35$. The arrival-time difference is calculated by $\Delta t = (t_{511,1}+t_{511,2})/2-t_\gamma$ where $t_{511,j}$'s are the measured arrival times of the annihilation photons, and $t_\gamma$ is the measured arrival time of the prompt gamma. The standard deviation in $\Delta t$ is therefore equal to  $\sqrt{3/2}\,\sigma_1=(\sqrt{3}/2)\,\sigma_\mathrm{TOF}$.}
    
    In theory, $\tilde{S}(\Delta t)$ decreases exponentially to zero as $\Delta t$ increases, which allows $\mu_n\{\tilde{S}(\Delta t)\}$ to be well defined. In reality, noise in $\tilde{S}(\Delta t)$ does not necessarily decrease with $\Delta t$ and hence will contribute a substantial statistical error in $\mu_n\{\tilde{S}(\Delta t)\}$, especially for large $n$. This instability can be considerably reduced by instead using $\mu_n\{e^{-s\Delta t}\tilde{S}(\Delta t)\}$, $s>0$ as the term $e^{-s\Delta t}$ attenuates the contribution from data at large $\Delta t$. Although a large $s$ is favored for alleviating the effects of noise, it also diminishes the differences among $s+\lambda_i$ and thereby requires a large $n$ for \EQ{\ref{eq:moment-att-PAL1}} to hold true. However, calculation of higher-order moments is more susceptible to noise so the value of $s$ needs to be chosen with care. In this work, in consideration of the numerical values of $\lambda_i$ and $I_i$, we choose $s=1$, which is empirically justified based on the results to be reported below. Future work will consider $s$ more extensively.

    From \EQ{\ref{eq:moment-defn}}, $\mu_n\{f(t)\}$ can be regarded as a filter that removes an increasingly wider range of small $t$ data as $n$ increases. By inspection of \FIG{\ref{fig:data}} it can be stated that by choosing an $n$ which makes \EQ{\ref{eq:moment-att-PAL1}} valid, a ``soft'' low cutoff has been introduced. This avoids using data where DA and p-Ps are dominant.
    
    \subsection{Simulated data}
\begin{figure}
            \centering
            \includegraphics[width=\linewidth]{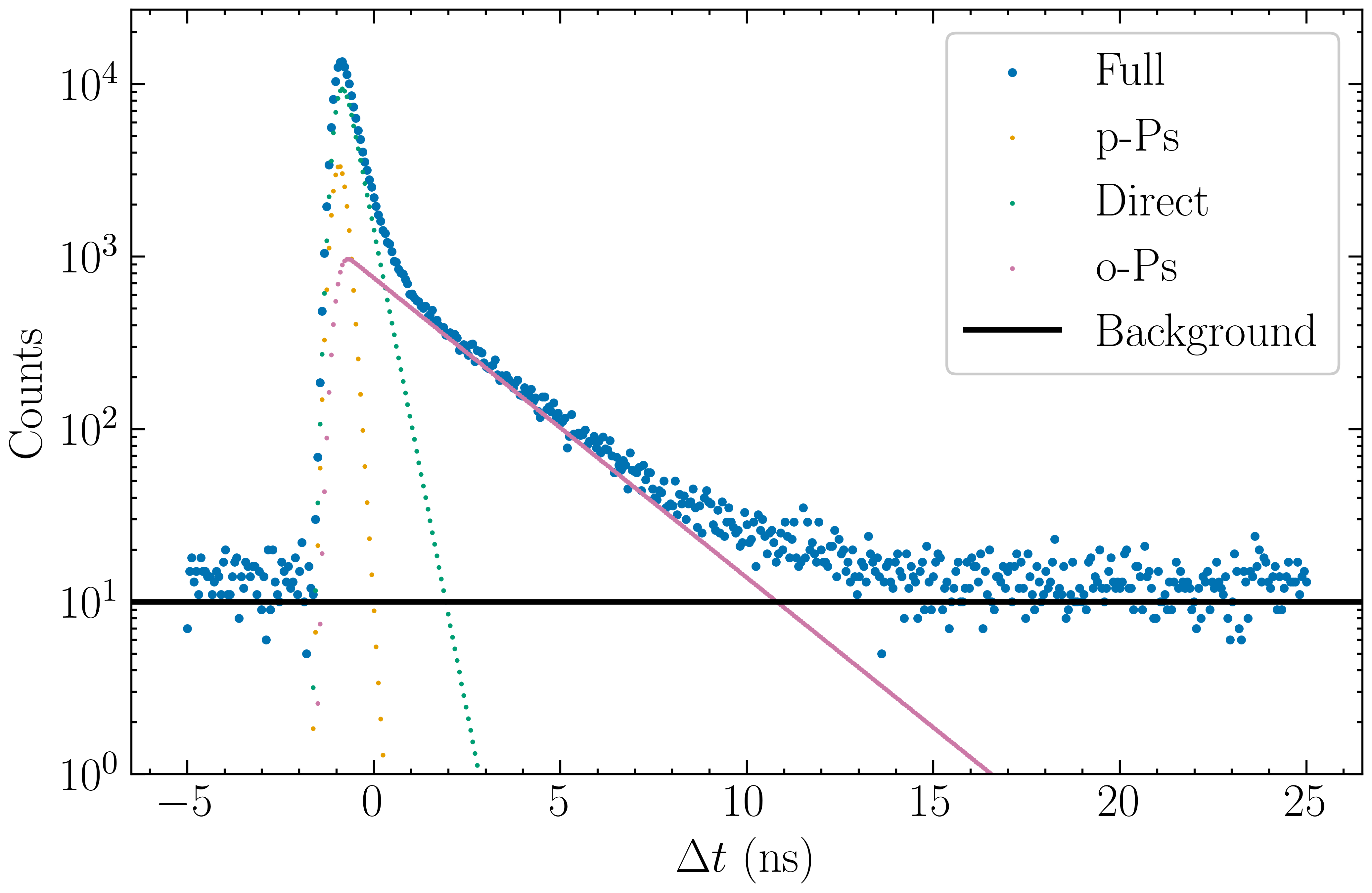}
            \caption{A simulated histogram of $\Delta t$ 
            having $\SI{2e5}{}$ events and
            using a \SI{60}{ps} bin size.
            It was generated by using the following parameters:
            $\lambda_1 I_1$, $\lambda_2 I_2$, $\lambda_3 I_3$  = $0.078$, $0.388$, $0.165$,
            $\tau_1$, $\tau_2$, $\tau_3 = \SI{0.125}{ns}$, $\SI{0.388}{ns}$, $\SI{2.5}{ns}$,
            $\sigma = \SI{0.169}{ns}$,
            $b = 10$,
            and $t_0 = \SI{9}{ns}$.
            $\lambda_1, \lambda_2, \lambda_3 =$  $\SI{8}{ns^{-1}}$, $\SI{2.58}{ns^{-1}}$, $\SI{0.4}{ns^{-1}}$. The individual contributions from the p-Ps, DA, and o-Ps components are shown in yellow, green, and magenta, respectively.}
            \label{fig:data}
        \end{figure}

    The proposed method is evaluated by using computer generated simulation data. All computation codes were implemented in Python version 3.11, using specified values for $b$, $I_i$, $\lambda_i$, and $\sigma$, and the desired total number of events for the histogram $N$. The simulation program first computed
    $S(\Delta t_i)$ according to \EQ{\ref{eq:signal}} at discrete times in [\SI{-5}{ns}, \SI{25}{ns}] and at a regular spacing of \SI{60}{ps}. Next, it scaled $S(\Delta t_i)$ to yield $\sum_i S(\Delta t_i)=N$. Then, the scaled $S(\Delta t_i)$ was replaced by an integer drawn by a Poisson random number generator (from the numpy.random module) whose mean equals the scaled $S(\Delta t_i)$. \FIG{\ref{fig:data}} shows an example of the generated histogram where the parameters were so chosen that it was similar to the measured histogram reported by Moskal et al.~\cite{moskal2021positronium}. Each simulated histogram, except where noted, contained $N=\SI{2e5}{}$ total events.

    \subsection{Implementation and numerical studies}
    \label{subsec:implementation}
    The background term $b$ was estimated using the average of the histogram in the $\Delta t < \SI{0}{ns}$ region. The estimate was then subtracted from each histogram data points and $\lambda_3$ was calculated according to \EQ{\ref{eq:moment-att-PAL1}} and \EQ{\ref{eq:estimate_lambda3}}. For computing moments, \EQ{\ref{eq:moment-defn}} was approximated by naive discrete summation: $\mu_n\{f(t)\} \approx  \delta t\sum_i t_i^n \tilde{S}(t_i)$ where $t_i$ are the time points where $f(t)$ is available and $\delta t$ is the spacing of $t_i$.
    
    The proposed method was evaluated for accuracy and precision against a number of parameters, including 
    1) the order of moment $n$ used,
    2) the upper truncation of the histogram data,
    3) the number of counts $N$, and
    4) the background level $b$.
    At present, the range of \textit{in vivo} o-Ps lifetime values has not been precisely established. However, in a recent paper Moskal et al. observed that
    cardiac myxoma and adipose tissue had mean lifetimes of \SI{1.912}{ns} and \SI{2.613}{ns}, respectively \cite{moskal2021positronium}. Therefore, we performed evaluations for o-Ps lifetimes of 1.0, 1.5, 2.0, 2.5, and 3.0 ns to cover the likely \textit{in vivo} lifetime range. On the other hand, since they are insensitive to the local environment \cite{cassidy2018experimental}, the reported mean values of \SI{388}{ps} and \SI{142}{ps} are used for DA and p-Ps lifetimes, respectively. For $I_1$, $I_2$ and $I_3$, the values used were based on quantitatively matching the simulated and measured data.

\section{Results}
\label{sec:results}
    The results are presented by \FIG{\ref{fig:estimate_vs_moment}} through \FIG{\ref{fig:estimate_vs_counts_sigma}}.  Each figure contains individual points which are estimates of $\tau_3$, denoted by $\hat{\tau}_3$ below, derived from our method. Each data point is the mean of the results obtained from \SI{1e4}{} histograms simulated by using the same parameters (\SI{1e5}{} in the case of \FIG{\ref{fig:estimate_vs_moment}}), and the shaded regions in the plots give the $\pm 1$ standard deviations (std) about the means. The horizontal lines, when present, indicate the true o-Ps lifetimes that are used to produce simulation data.
    
    \subsection{Lifetime estimate versus order of moment \texorpdfstring{$n$}{}} \label{subsec:leftime_vs_moment} 
\begin{figure}[tb]
        \centering
        \includegraphics[width=\linewidth]{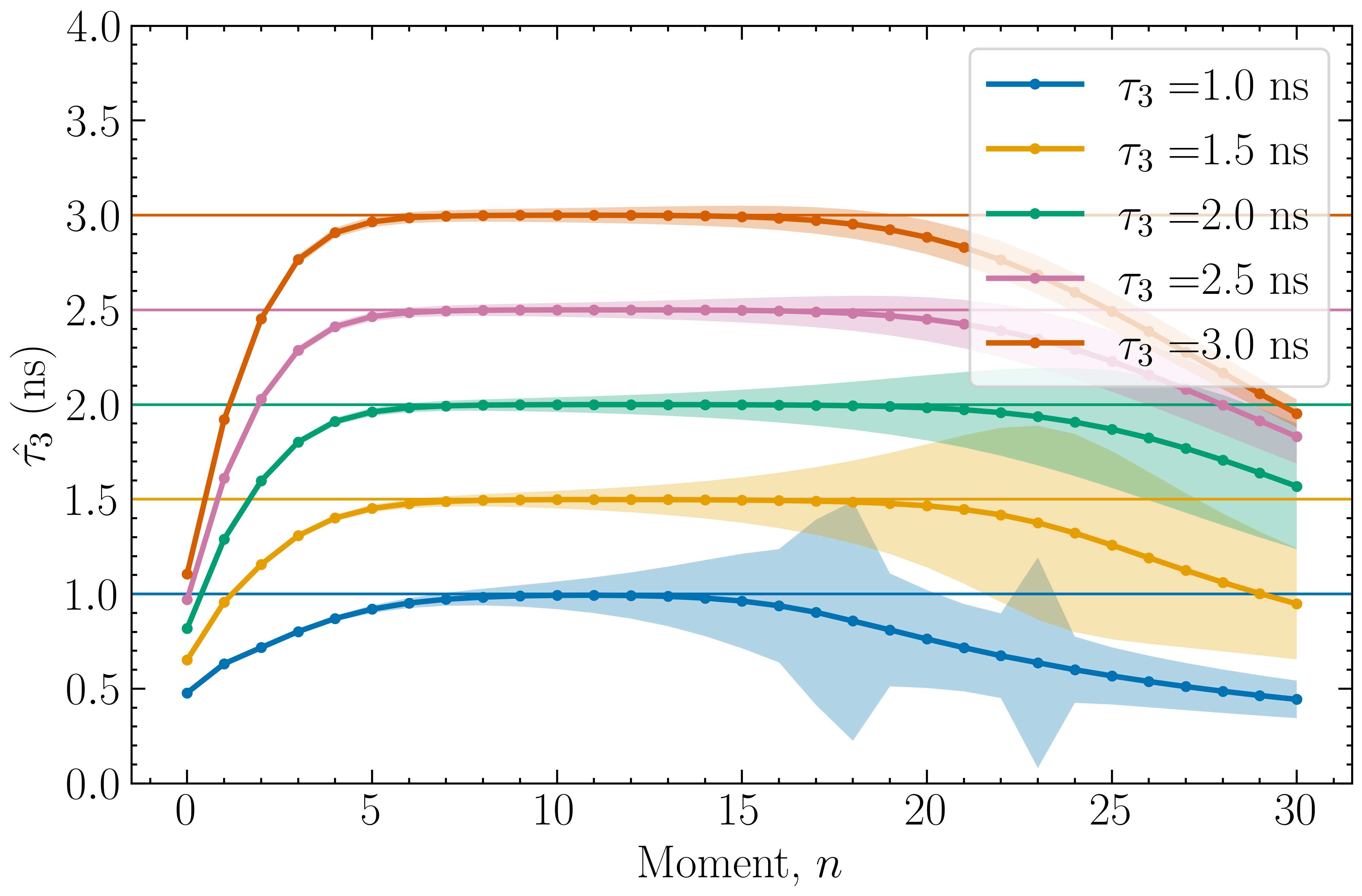}
        \caption{Estimated o-Ps lifetime as a function of the order of moment $n$. The shaded areas for each curve represent the $\pm 1$ standard deviation on each data point. The true $\tau_3$ values shown in the legend are those used for producing simulation data.}
        \label{fig:estimate_vs_moment}
    \end{figure}

\FIG{\ref{fig:estimate_vs_moment}} shows the estimated o-Ps lifetime when the order of moment $n$  employed by \EQ{\ref{eq:estimate_lambda3}} is varied. Four general trends can be observed.
    First, all curves show a plateau where the estimated lifetime has essentially zero bias. This plateau occurs between $n\approx5$ and $n\approx16$ depending on the o-Ps lifetime. The standard deviations of the estimates are also sufficiently small to allow for discrimination of all the lifetimes examined. Second, the standard deviation increases with $n$, which is consistent with the observation made above that higher-order moments are more sensitive to data noise. Third, as $\tau_3$ increases the plateau occurs at a lower $n$. This is because the differences between $s+\lambda_3$ and $s+\lambda_i$, $i=1,2$ increases, allowing the $i=3$ term to dominate the sum in \EQ{\ref{eq:moment-att-PAL}} at a smaller $n$. Fourth, all curves decrease toward zero as the order increases. This is because higher-order moments are increasingly more contributed by data at larger $\Delta t$ while data is simulated only for $-\SI{5}{ns}\leq\Delta t\leq\SI{25}{ns}$. In practice, the measured histogram is necessarily truncated.
    
    It is also noted that all curves converge at small $n$. This reflects the situation that with a small $n$ the $i=1,2$ terms in fact dominate the $i=3$ term in \EQ{\ref{eq:moment-att-PAL}}. As a result, \EQ{\ref{eq:estimate_lambda3}} yields some fixed value since $I_{1,2}$, and $\tau_{1,2}$ are constants. Based on the plot, we propose to that in general $n\lesssim 5$ should not be used for PALS o-Ps estimations. Table~\ref{tab:summary} summarizes the bias-minimizing order of moment in the plateau region and the statistics of the estimates obtained when using this order. As shown, the largest std/mean ratio is only 0.654\%, obtained for $\tau_3=\SI{1.0}{ns}$.

    \bgroup
        \def\arraystretch{\myrowsep}
        \setlength{\tabcolsep}{4.3pt}
        \begin{table}[tb] 
        \mytablefontsize
        \captionsetup{style=mystyle}
        \captionof{table}{Bias-minimizing estimates of Figure \ref{fig:estimate_vs_moment}, shown with the moment $n$ which minimized the bias. We again emphasize that this moment-based method requires $n$ to be sufficiently large.}
        \centering
        \begin{tabular}{lcccc
        >{\columncolor[HTML]{FFFFFF}}c }
        \hline
        \multirow{2}{*}{$\tau_3$ (ns)} & \multirow{2}{*}{order $n$} & \multicolumn{3}{c}{$\hat{\tau}_3$} \\\cline{3-6}
        \ & \ & mean (ns) & std (ns) & std/mean & \% error \ \\
        \hline\hline
        1.0  & 11 & 0.99 & 0.094 & 0.095 & 0.654 \\
        1.5  & 11 & 1.50 & 0.056 & 0.037 & 0.096 \\
        2.0  & 12 & 2.00 & 0.051 & 0.025 & 0.008 \\
        2.5  & 10 & 2.50 & 0.041 & 0.016 & 0.001\\
        3.0  & 10 & 3.00 & 0.037 & 0.012 & 0.009 \\
        \hline\hline

        \end{tabular}%
        \label{tab:summary}
        \end{table}
    \egroup 
    
\begin{figure}[tb]
        \centering
        \includegraphics[width=\linewidth]{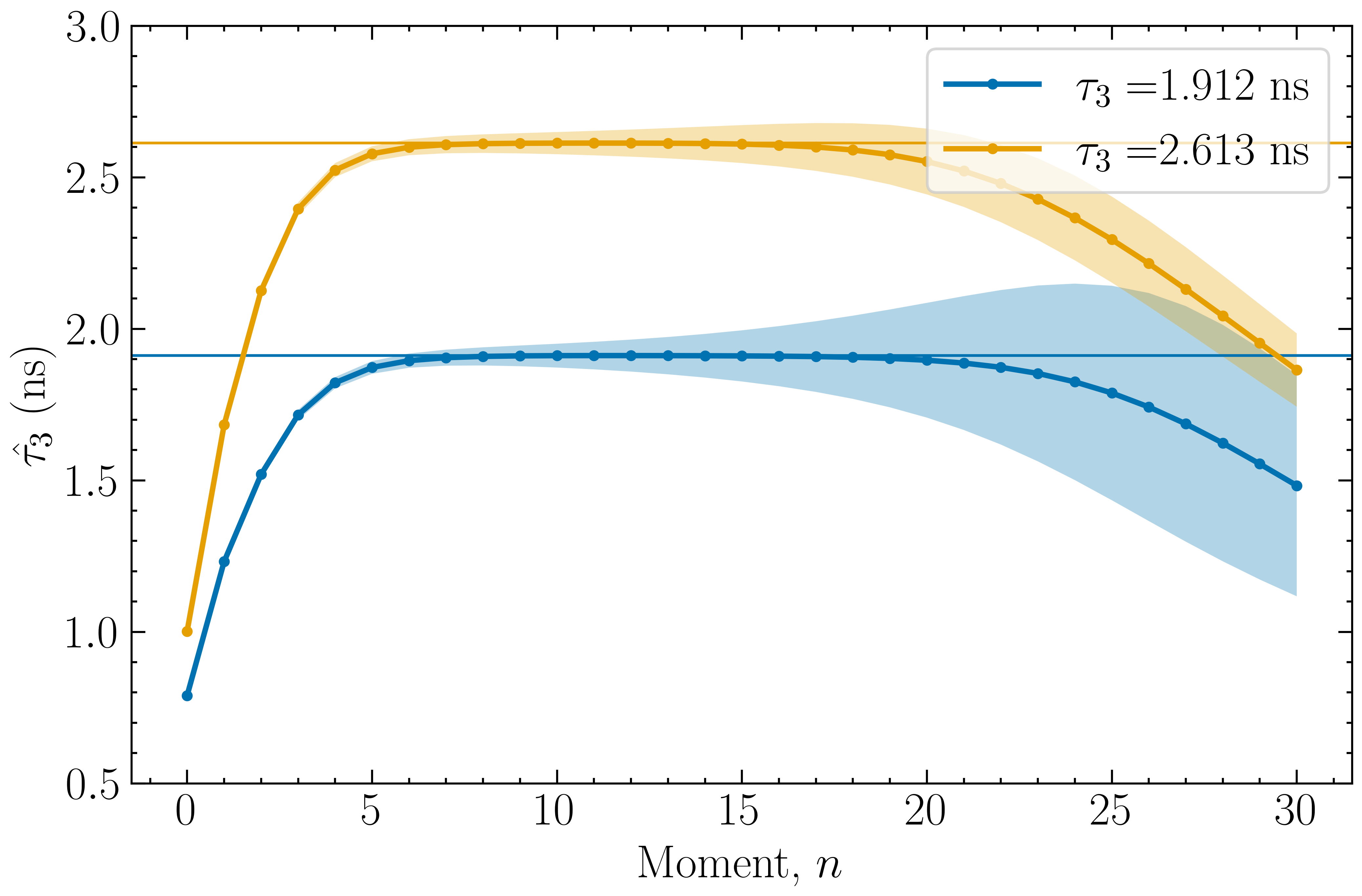}
        \caption{Similar to Figure \ref{fig:estimate_vs_moment}, this shows two practical cases for o-Ps lifetimes which have been measured recently \textit{in vivo} \cite{moskal2021positronium}.} 
        \label{fig:estimate_vs_moment_moskal}
    \end{figure}

    Figure \ref{fig:estimate_vs_moment_moskal} shows the result obtained when the $\tau_3$ values measured for cardiac myxoma and adipose tissue by Moskal et al. \cite{moskal2021positronium} were used to produce simulation data. In this case, the bias-minimizing orders of the moments were found to be $n=11$ for $\tau_3=\SI{1.912}{ns}$ (myxoma tissue), which yielded $\hat{\tau}_3 = 1.91 \pm 0.05$ ns, and $n=11$ for $\tau_3=\SI{2.613}{ns}$ (adipose tissue), which yielded $\hat{\tau}_3 = 2.61 \pm 0.04$ ns. The means of these estimates are within $0.02\%$ and $0.005\%$ of their respective true values.
    
    \subsection{Lifetime estimate versus number of events \texorpdfstring{$N$}{}}
    \label{subsec:lifetime_vs_event_number}

\begin{figure}[tb]
        \centering
        \includegraphics[width=\linewidth]{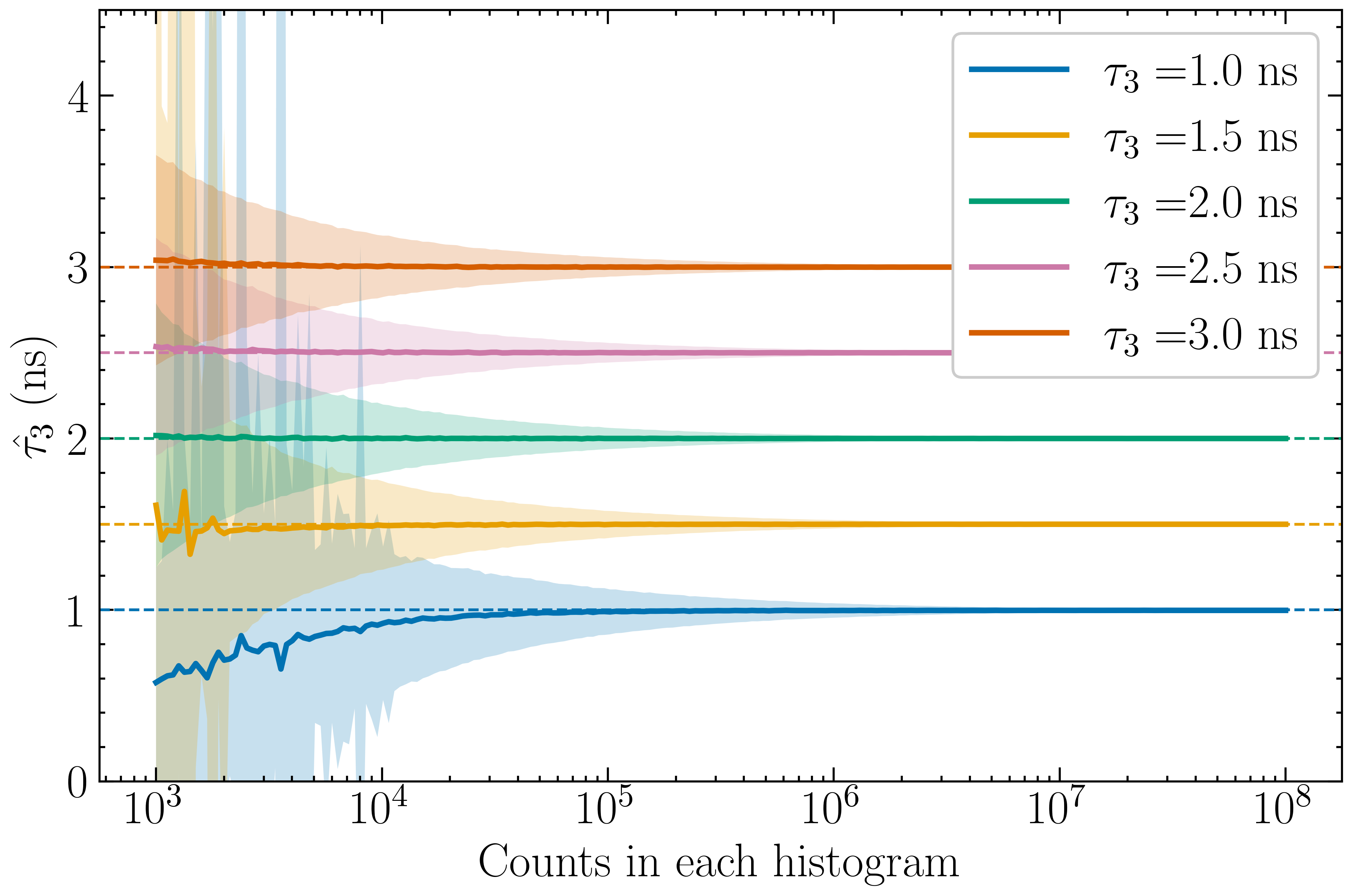}
        \caption{Estimated o-Ps lifetimes as a function
        of the number of events $N$ in the histogram. 
        The shaded areas again indicate $\pm 1$ standard deviation about the mean.}
        \label{fig:estimate_vs_counts}
    \end{figure}

    The results discussed were obtained by using simulated data with $N=\SI{2e5}{}$. Generally, a histogram derived from a larger number of events has better statistics and can lead to better estimates for $\lambda_3$. Figure \ref{fig:estimate_vs_counts} shows the results for $N$ ranging from \SI{1e3}{} and \SI{1e8}{}. The order of moment was fixed to $n=11$. A tabulated summary of these results are shown in Table \ref{tab:counts}.
    
    \bgroup
        \def\arraystretch{\myrowsep}
        \setlength{\tabcolsep}{5pt}
        \begin{table}[tb] 
        \mytablefontsize
        \captionsetup{style=mystyle}
        \captionof{table}{Selected estimates from \FIG{\ref{fig:estimate_vs_counts}}.}
        \centering
        \begin{tabular}{cccccc
        >{\columncolor[HTML]{FFFFF}}c }
        \hline
        \multirow{2}{*}{$\tau_3$ (ns)} & \multicolumn{4}{c}{$\hat{\tau}_3$ (ns)} \\\cline{2-5}
         &\multicolumn{1}{c}{$N=\SI{e3}{}$} & \multicolumn{1}{c}{$N=\SI{e4}{}$} & \multicolumn{1}{c}{$N=\SI{e5}{}$} & \multicolumn{1}{c}{$N=\SI{e6}{}$}\\
        \hline\hline
        1.0  & $0.6 \pm 0.6$ & $0.9 \pm 0.4$ & $1.0 \pm 0.1$ & $0.997 \pm 0.004$ \\
        1.5  & $2 \pm 16$ & $1.4 \pm 0.3$ & $1.49 \pm 0.08$ & $1.498 \pm 0.002$  \\
        2.0  & $2.0 \pm 0.8$ & $2.0 \pm 0.2$ & $2.00 \pm 0.06$ & $2.000 \pm 0.002$ \\
        2.5  & $2.5 \pm 0.6$ & $2.5 \pm 0.2$ & $2.50 \pm 0.06$ & $2.500 \pm 0.001$ \\
        3.0  & $3.0 \pm 0.6$ & $3.0 \pm 0.1$ & $3.00 \pm 0.06$ & $3.000 \pm 0.001$ \\
        \hline\hline

        \end{tabular}%
        \label{tab:counts}
        \end{table}
    \egroup 

\begin{figure}[tb]
        \centering
        \includegraphics[width=\linewidth]{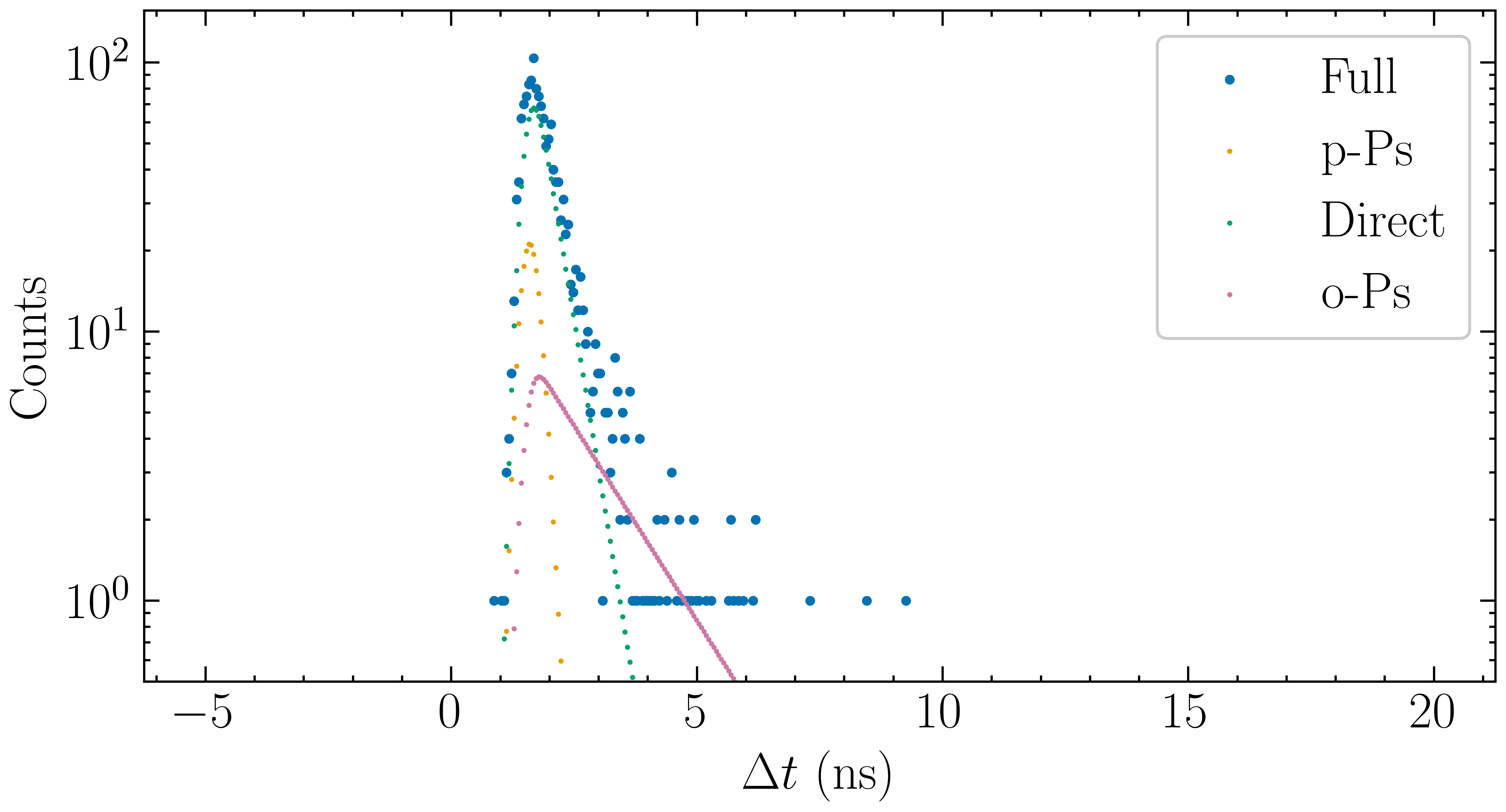}
        \caption{A simulated histogram for
        $\tau_3 = \SI{1.5}{ns}$ and $N=1,500$,
        including the noise-free contributions
        of DA, p-Ps, and o-Ps.}
        \label{fig:low_stat}
    \end{figure}

    It can be seen that, as expected, for all the $\tau_3$ examined the standard deviations of the estimates decrease continually as $N$ increases. The estimates asymptotically approach the true values as $N$ increases and the statistics are seen to improve. The occurrences of some exceptionally large standard deviations reflect the instability of the ratio estimate given by  \EQ{\ref{eq:estimate_lambda3}} when the denominator is erroneously small with respect to the numerator due to data noise. Figure~\ref{fig:low_stat} shows a simulated histogram for $N=\SI{1.5e3}{}$ and $\tau_3 = \SI{1.5}{ns}$ when one such case occurs. It can be seen that the $\Delta t > 4$ region of the histogram where o-Ps is assumed to dominate is sparsely populated, leading to large and inconsistent errors in  the numerator and denominator of \EQ{\ref{eq:estimate_lambda3}}. Controlling this variability will be commented on in \SECT{\ref{drawbacks_and_future}}.

    \subsection{Lifetime estimate versus cutoff and background}
    \label{subsec:lifetime_vs_cutoff}

\begin{figure}[tb]
        \centering
        \includegraphics[width=\linewidth]{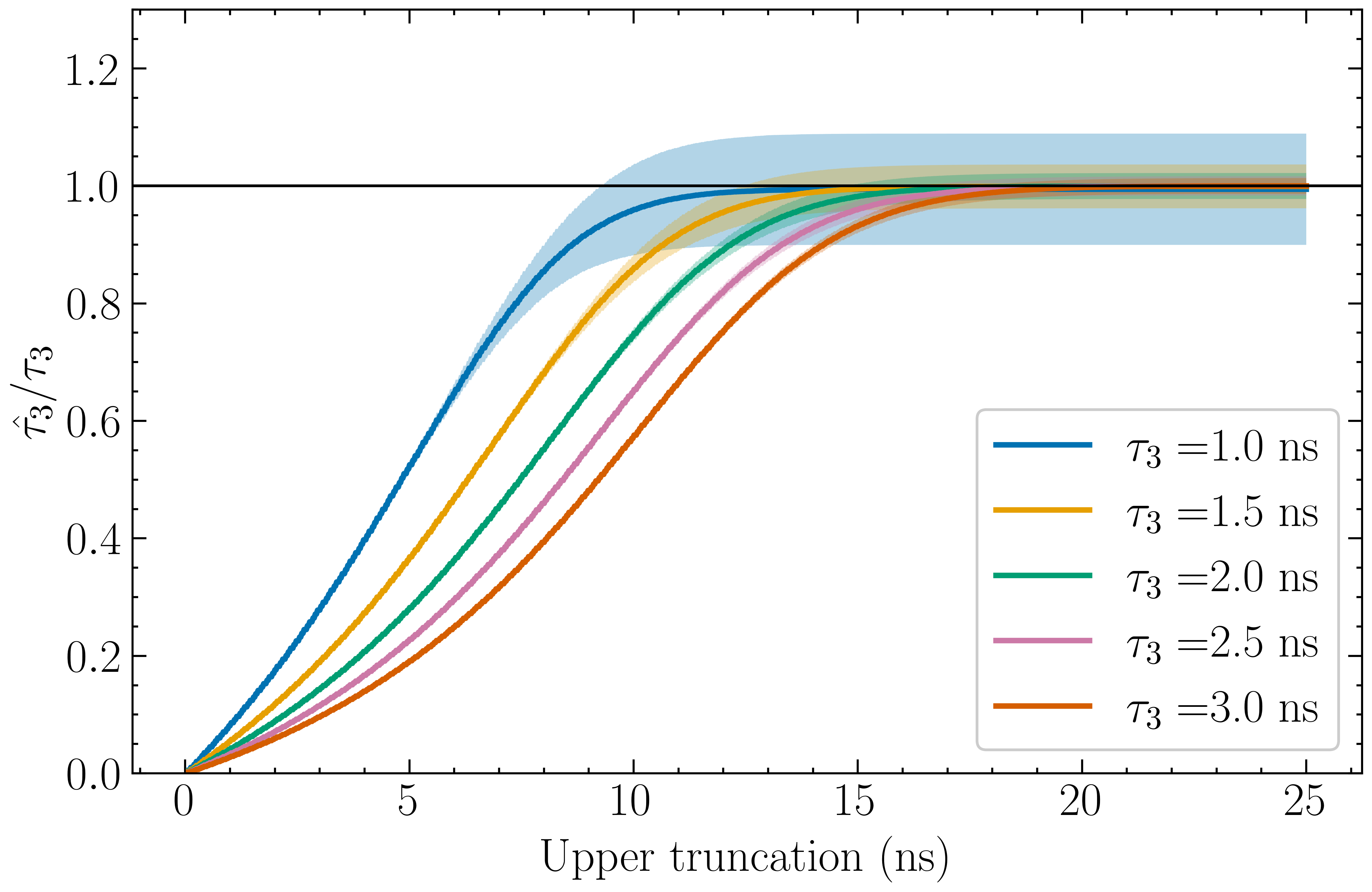}
        \caption{Estimated o-Ps lifetime as a function of the upper truncation threshold $\Delta t_{up}$. Note that here $\hat{\tau}_3/\tau_3$ is plotted.
        The horizontal line indicates the perfect estimate given by $\hat{\tau}_3/\tau_3 = 1$.}
        \label{fig:upper_truncation}
    \end{figure}

\begin{figure}[tb]
        \centering
        \includegraphics[width=\linewidth]{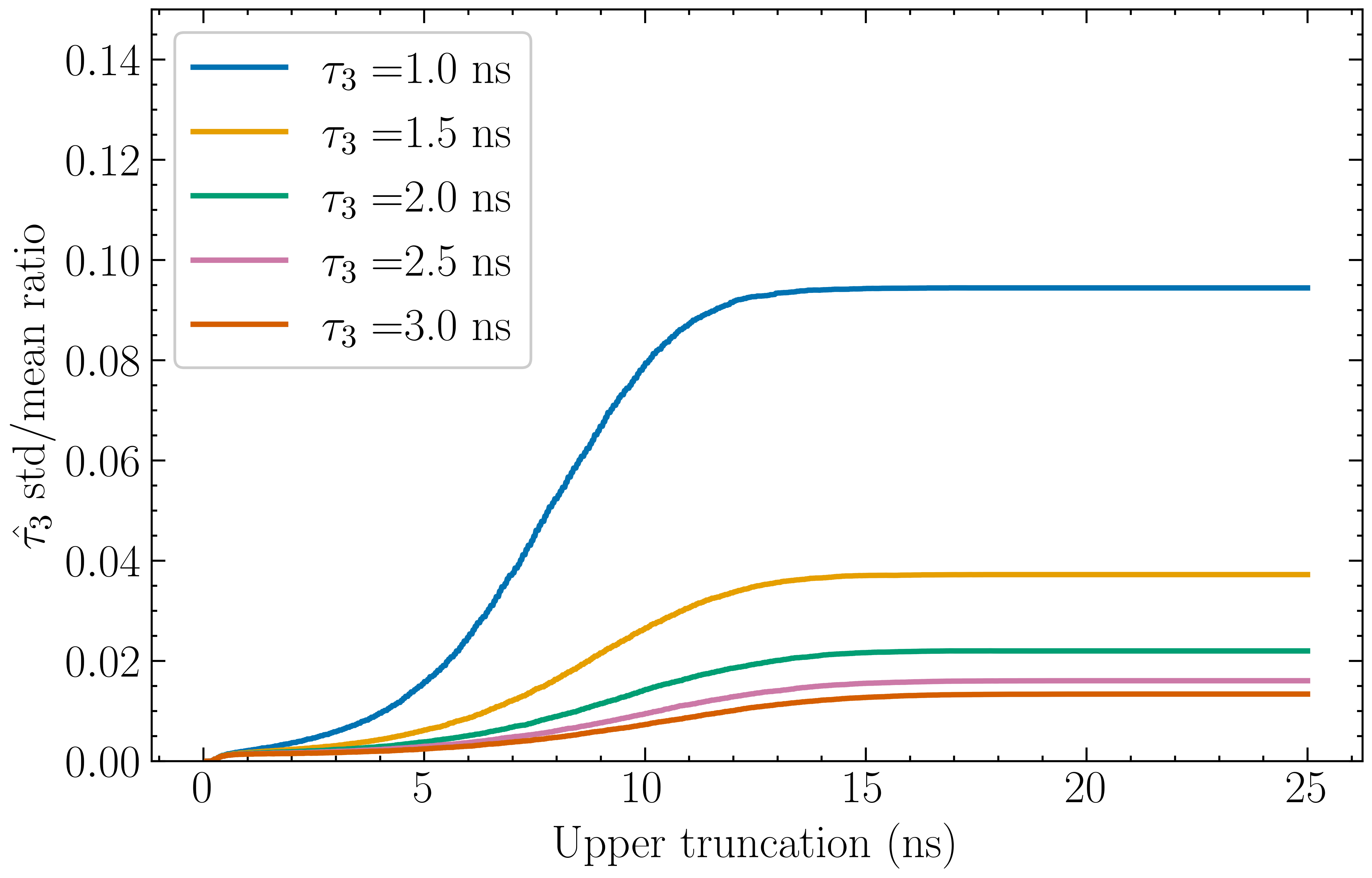}
        \caption{The ratio of the standard deviation to the mean of the curves from Figure \ref{fig:upper_truncation}. The ratios are seen to decrease as the upper truncation decreases. The standard deviations about the mean values are too small to be visible.}
        \label{fig:ratio_upper_truncation}
    \end{figure}

    In an attempt to further alleviate the deleterious effects of noise, we also examined removing data above certain $\Delta t$. Figures \ref{fig:upper_truncation} shows that, for the case of $n=11$, as the upper truncation threshold $\Delta t_{up}$ is decreased, the standard deviation decreases, which is particularly evident for small $\tau_3$, whereas the bias increases in the form of underestimation. However, as shown in Figure \ref{fig:ratio_upper_truncation}, the standard deviations relative to the distances between means increases as $\Delta t_{up}$ is lowered, at least for $\Delta t_{up}$ above $\sim\SI{10}{ns}$. Therefore, at least for discrimination tasks, it is beneficial to apply truncation even though it introduces bias.
    
\begin{figure}[tb]
        \centering
        \includegraphics[width=\linewidth]{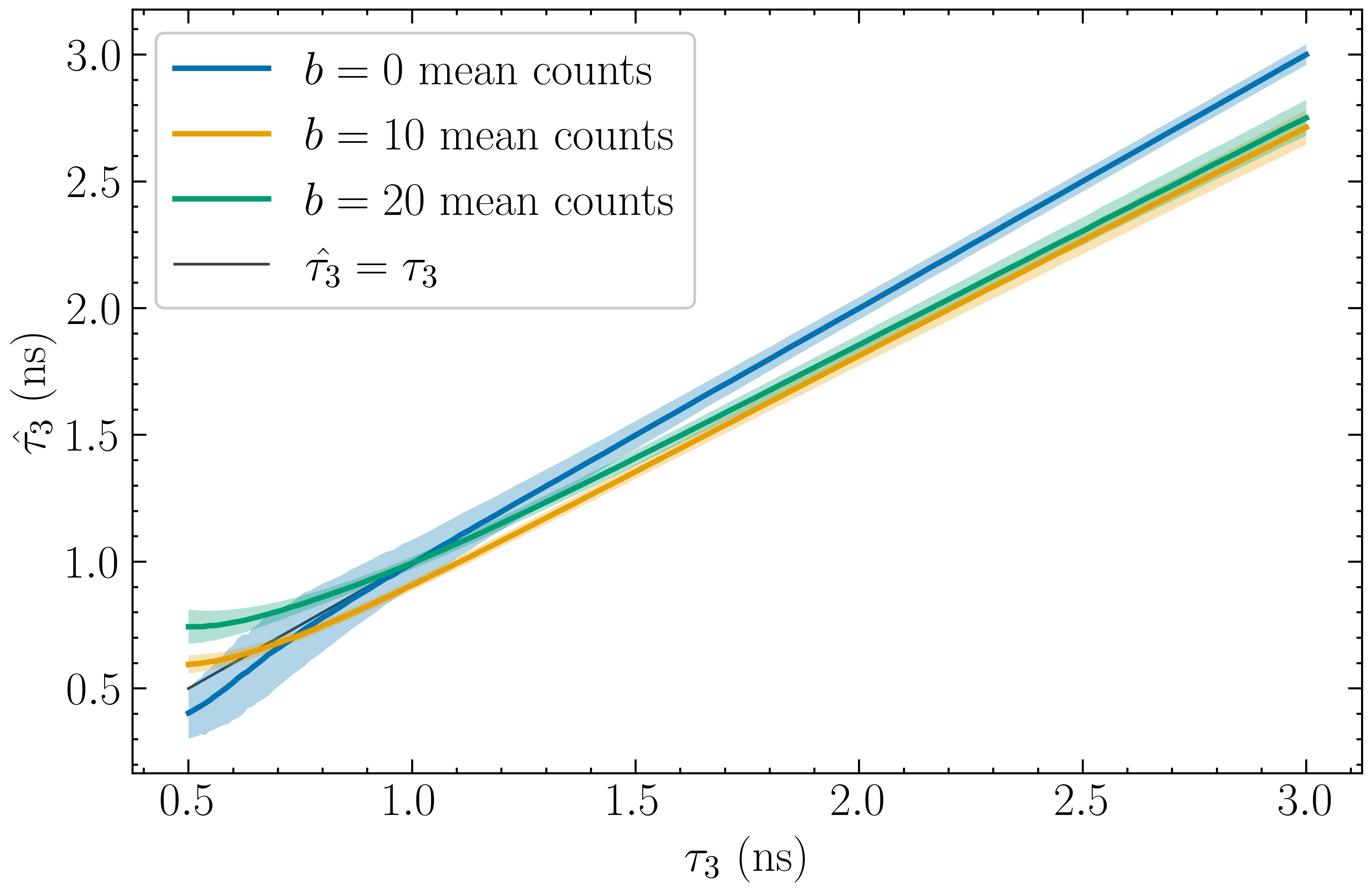}
        \caption{Estimated o-Ps lifetime versus the true value for three background levels $b$ with $n=11$, $N=\SI{2e5}{}$ and no upper truncation. The region close to $\tau_3 = \SI{0.5}{ns}$ may not be monotonic for some cases due to the curvature.}
        \label{fig:monotonic}
    \end{figure}

    The presence of a nonzero background $b$ increases data noise in two manners.
    First, for a given number of total events the number of events contributing to the signal is lowered. Second, although $b$ may be estimated and subtracted from  $S(\Delta t)$, the noise associated with it remains in the data. Due to its Poisson nature, the variance of the associated noise is proportional to $b$. \FIG{\ref{fig:monotonic}} shows the relationship between the estimate $\hat{\tau}_3$ and the true $\tau_3$ for three background levels with $n=11$, $N=\SI{2e5}{}$, and no upper truncation. The background is modeled as Poisson noise with means of $b=0, 10,$ and $20$  counts.
    
    \subsection{Monotonicity}
    \label{monotonic_section}    

    We have observed that for discrimination tasks it is beneficial to allow for for some bias if the std/mean ratio of the estimate decreases. For quantitative tasks, as long as $\hat{\tau}_3$  is related to $\tau_3$ monotonically in the mean, this observation remains true because we can correct for the bias in $\hat{\tau}_3$ by using, for example, a predetermined calibration curve. Such monotonicity is observed in both \FIG{\ref{fig:upper_truncation}} and \FIG{\ref{fig:monotonic}}. Since curves in \FIG{\ref{fig:estimate_vs_moment}} do not cross one another this monotonicity is also true when varying $n$. For $b=0, 10,$ and $20$ mean counts, $\hat{\tau}_3$ remains monotonic with $\tau_3$ in the mean and is approximately linear for $\tau_3\gtrsim\SI{1.5}{ns}$. Below $\sim\SI{0.5}{ns}$, monotonicity may not hold for $n = 11$, however this is well below the range for currently-known biological values of $\tau_3$.

\section{Discussion}
    \subsection{Effect of the number of events}
    It was observed that our moment-based method is significantly affected by the presence of Poisson noise, which comes with reduced event numbers. However, in the mean this  method is still monotonic with the true value. A clinical PET system would produce a single histogram for each voxel (even though this is not necessary for estimating $\tau_3$ using our method). A typical scan may consist of approximately $\SI{185}{MBq}$ $(\SI{5}{mCi})$ of injected activity and a \SI{64}{mm^3} voxel size (\SI{4}{mm} side length). Assuming the human body to be mainly composed of water with a total mass of $\SI{80}{kg}$, it would have approximately $\SI{8e4}{cm^3}$ of internal volume. The activity in this volume would then be \SI{2.3e3}{Bq/cm^3} and we might expect \SI{148}{Bq/voxel} if there is a uniform activity distribution. Assuming a per-voxel sensitivity of $1\%$, 1.48 counts would be collected each second. To obtain \SI{e3}{} counts for each voxel, we would need to collect data for approximately 11.3 minutes. For reference, Moskal et al. reported human brain PLI data, where the scan time was 10 minutes (after a standard radiotracer distribution waiting period), and the number of counts collected were 342 for the healthy brain tissue, 547 for the tumor, and 1119 for the salivary glands \cite{moskal2024first}.
    
    Additionally, above we have assumed a uniform radiotracer distribution. When using tumor-specific radiotracers, the radioactivity can be concentrated to tumor regions which will improve the statistics.


    \subsection{Short o-Ps lifetimes \textit{in vivo}} \label{disc1}
    Thus far, the shortest published o-Ps lifetime value using a phantom was $\SI{1.8239}{ns}$ \cite{shibuya2020oxygen}. This was measured in fully $\mathrm{O}_2$ saturated water. In a more artificial environment, Stepanov et al. \cite{stepanov2020dissolved} bubbled oxygen gas through water and measured an o-Ps lifetime of \SI{1.746}{ns}. With argon gas bubbled through water, however, a lifetime of \SI{1.833}{ns} was measured. This is evidence that even extremely well-oxygenated \textit{in vivo} environments will not have o-Ps lifetimes of less than \SI{1.746}{ns}. In a recent review, Moskal and St\c{e}pie\'{n} concluded that the mean biological o-Ps lifetime would be approximately \SI{2}{ns} \cite{moskal2021positronium_review}. 
    
    This moment-based method has been shown to produce viable results for lifetimes above approximately \SI{1}{ns}, depending on noise, moment, and truncation. For lifetimes on that are on the order of \SI{1}{ns}, standard deviations for $n>15$ (seen in Figure \ref{fig:estimate_vs_moment}) grow rapidly, and estimations may no longer be monotonic.

    \subsection{Drawbacks and future work} \label{drawbacks_and_future}

    A main drawback of this method is the response to noise. The standard deviations in Figure \ref{fig:estimate_vs_counts} are a significant fraction of the estimate itself, even though the mean is stable. However, this method will be computationally less expensive to execute over an entire PET dataset than fitting-based methods, which may outweigh its precision for low-count cases. Shibuya et al. has previously calculated that approximately \SI{3e8}{counts} will be needed to estimate oxygen concentration to a precision of \SI{10}{mmHg} \cite{shibuya2020oxygen}. This would correspond to three times as many counts as the extent of Figure \ref{fig:estimate_vs_counts}. At this point, the method has essentially reached its asymptote, minimum bias, and minimum standard deviation.
 
    \begin{figure}[tb]
        \centering
        \includegraphics[width=\linewidth]{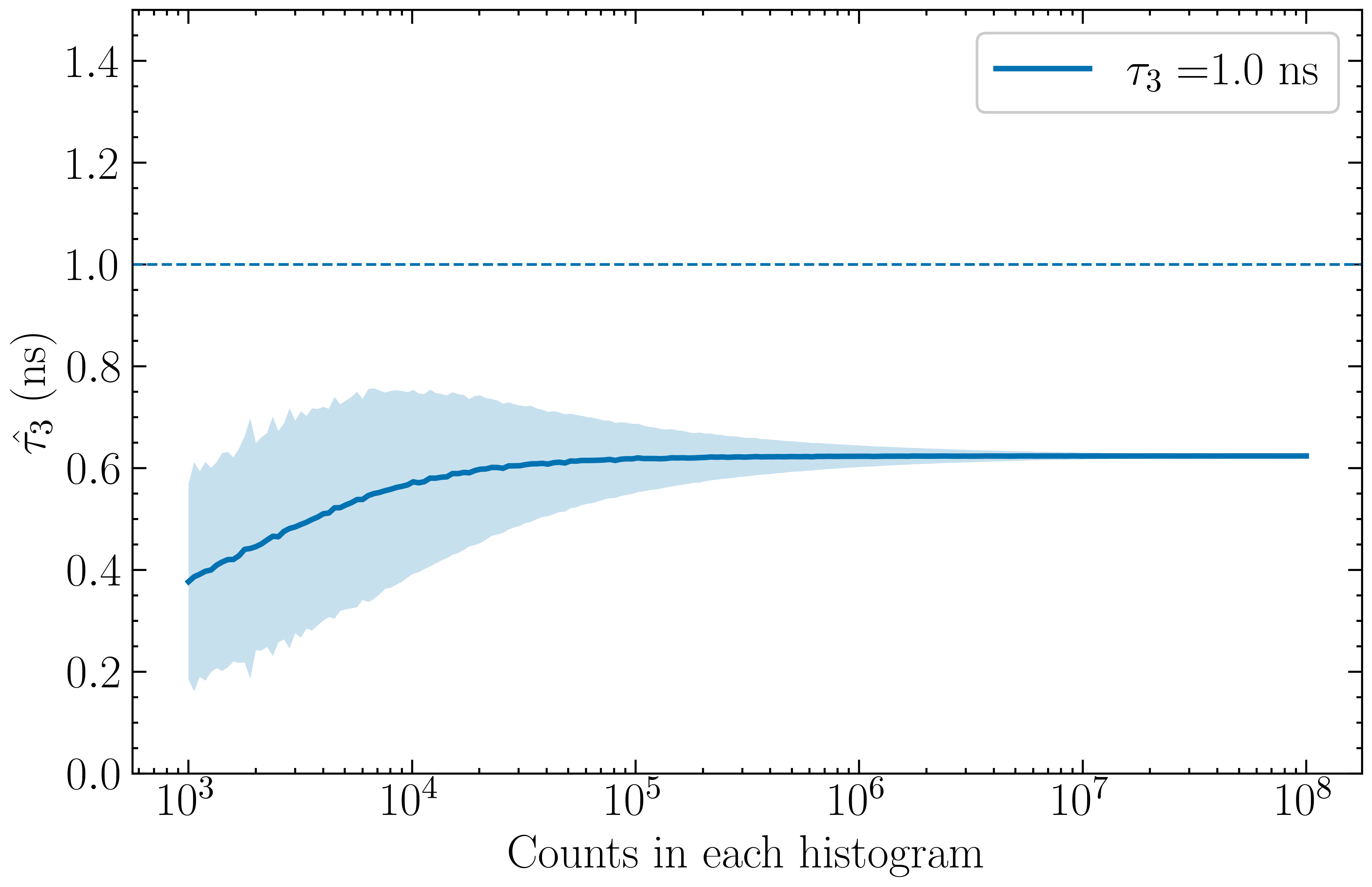}
        \caption{The estimate of the $\tau_3 = \SI{1}{ns}$ case from Figure \ref{fig:estimate_vs_counts}, however, with $\delta = 0.3$ to demonstrate how the standard deviation variability can be handled for small $\tau_3$ values.}
        \label{fig:estimate_vs_counts_sigma}
    \end{figure}
    
    To control variations in the standard deviation at low counts, a small positive number $\delta$ may be introduced, as shown in Equation \ref{eq:estimate_lambda3}. This results in an increased bias and decreased standard deviation seen in Figure \ref{fig:estimate_vs_counts_sigma}. Despite the increased overall bias in the asymptotic region (37.6\% error), the curve remains monotonic and asymptotic, allowing for the bias to be easily corrected. For comparison, with \SI{1e3} histogram counts, the standard deviation decreased from \SI{0.62}{ns} for $\delta = 0$ to \SI{0.19}{ns} for $\delta = 0.3$, while the percentage error increased from 42.9\% to 62.4\% for the two cases, respectively. 
    
    Future work on this estimation method will include its testing on full three-dimensional reconstruction, testing its computational speed with respect to current estimation techniques, and further optimization of fitting parameters, such as the lower truncation point, the $e^{-st}$ term, smoothing of the histogram, and noise handling modifications.

\section{Conclusion}
In this report we present an analytical method to estimate the orthopositronium lifetime in PALS measurements. This method uses moments of the histogram of arrival time differences, and employs an exponential weighting to mitigate numerical instability in calculation of moments from noisy data. The moment-based method was characterized in this work, and it was shown to be a stable, monotonic estimate in most cases. For cases in which the standard deviation was large, modifications to the method may be employed. This method will continue development to control noise for cases with small statistical power, and will be implemented and tested for three-dimensional PET images. 

\section*{Appendix}

From $f_n(\lambda)=\mu_n\{\mathrm{EXP}(t;\lambda)\}=\lambda \int_0^\infty dt\ t^n e^{-\lambda t}$,
one can immediately derive
\begin{equation}
    \frac{df_n(\lambda)}{d\lambda} = \frac{1}{\lambda}f_n(\lambda) - f_{n+1}(\lambda).
\end{equation}
Using Equation~\ref{eq:moment-EXP} in the above equation then yields
\begin{equation}
    f_{n+1}(\lambda)
    = \frac{1}{\lambda}\frac{n!}{\lambda^n} + n\frac{n!}{\lambda^{n+1}} 
    = \frac{(n+1)!}{\lambda^{n+1}}.
\end{equation}
When $n=0$, Equation~\ref{eq:moment-EXP} also correctly yields $f_0(\lambda)=1$.
Therefore, by induction Equation~\ref{eq:moment-EXP} is valid for $n\geq 0$.

The function $g(t;\lambda,\sigma)=\mathrm{EXP}(t;\lambda)\ast\mathcal{N}(t;\sigma)$ is known as the Exponential Modified Gaussian (EMG) and can be shown to equal 
\begin{equation}
g(t;\lambda,\sigma) = \lambda\ e^{-\lambda t}\ e^{\sigma^2\lambda^2/2}\ h(t;\lambda,\sigma),
\end{equation}
where
\begin{equation}
    h(t;\lambda,\sigma) = \frac{1}{2} \left(
    1 + \mathrm{erf}\left( \frac{t-\sigma^2 \lambda}{\sqrt{2}\sigma} \right)
    \right)
\end{equation}
and $\mathrm{erf}(t)$ is the error function \cite{hanggi1985errors}.
At large $t$, $h(t;\lambda,\sigma)\approx 1$ because $\mathrm{erf}(t)\approx 1$.
Therefore,
\begin{equation}
\label{eq:appendix-EMG-approx}
e^{-st}g(t;\lambda,\sigma)
\approx\frac{\lambda}{s+\lambda}\ e^{\sigma^2\lambda^2/2}\ \mathrm{EXP}(t;s+\lambda).
\end{equation}
Note that, in evaluating $\mu_n\{f(t)\} = \int dt\ t^n f(t)$ the term $t^n$ in the integral
progressively diminishes the contribution of $f(t)$ at small $t$ as $n$ increases.
Therefore, for sufficiently large $n$ one can use the approximation 
given by Equation~\ref{eq:appendix-EMG-approx} to 
evaluate $\mu_n\{e^{-st}g(t;\lambda,\sigma)\}$,
which then yields Equation~\ref{eq:moment-EMG}.

\bibliography{bib}

\begin{thebibliography}{10}%
\makeatletter
\providecommand \@ifxundefined [1]{%
 \@ifx{#1\undefined}
}%
\providecommand \@ifnum [1]{%
 \ifnum #1\expandafter \@firstoftwo
 \else \expandafter \@secondoftwo
 \fi
}%
\providecommand \@ifx [1]{%
 \ifx #1\expandafter \@firstoftwo
 \else \expandafter \@secondoftwo
 \fi
}%
\providecommand \natexlab [1]{#1}%
\providecommand \enquote  [1]{``#1''}%
\providecommand \bibnamefont  [1]{#1}%
\providecommand \bibfnamefont [1]{#1}%
\providecommand \citenamefont [1]{#1}%
\providecommand \href@noop [0]{\@secondoftwo}%
\providecommand \href [0]{\begingroup \@sanitize@url \@href}%
\providecommand \@href[1]{\@@startlink{#1}\@@href}%
\providecommand \@@href[1]{\endgroup#1\@@endlink}%
\providecommand \@sanitize@url [0]{\catcode `\\12\catcode `\$12\catcode `\&12\catcode `\#12\catcode `\^12\catcode `\_12\catcode `\%12\relax}%
\providecommand \@@startlink[1]{}%
\providecommand \@@endlink[0]{}%
\providecommand \url  [0]{\begingroup\@sanitize@url \@url }%
\providecommand \@url [1]{\endgroup\@href {#1}{\urlprefix }}%
\providecommand \urlprefix  [0]{URL }%
\providecommand \Eprint [0]{\href }%
\providecommand \doibase [0]{https://doi.org/}%
\providecommand \selectlanguage [0]{\@gobble}%
\providecommand \bibinfo  [0]{\@secondoftwo}%
\providecommand \bibfield  [0]{\@secondoftwo}%
\providecommand \translation [1]{[#1]}%
\providecommand \BibitemOpen [0]{}%
\providecommand \bibitemStop [0]{}%
\providecommand \bibitemNoStop [0]{.\EOS\space}%
\providecommand \EOS [0]{\spacefactor3000\relax}%
\providecommand \BibitemShut  [1]{\csname bibitem#1\endcsname}%
\let\auto@bib@innerbib\@empty
\bibitem [{\citenamefont {Moskal}\ and\ \citenamefont {St{\k{e}}pie{\'n}}(2022)}]{moskal2022positronium_review}%
  \BibitemOpen
  \bibfield  {author} {\bibinfo {author} {\bibfnamefont {P.}~\bibnamefont {Moskal}}\ and\ \bibinfo {author} {\bibfnamefont {E.~{\L}.}\ \bibnamefont {St{\k{e}}pie{\'n}}},\ }\bibfield  {title} {\bibinfo {title} {Positronium as a biomarker of hypoxia},\ }\href@noop {} {\bibfield  {journal} {\bibinfo  {journal} {Bio-Algorithms and Med-Systems}\ }\textbf {\bibinfo {volume} {17}},\ \bibinfo {pages} {311} (\bibinfo {year} {2022})}\BibitemShut {NoStop}%
\bibitem [{\citenamefont {Moskal}\ \emph {et~al.}(2021)\citenamefont {Moskal}, \citenamefont {Dulski}, \citenamefont {Chug}, \citenamefont {Curceanu}, \citenamefont {Czerwi{\'n}ski}, \citenamefont {Dadgar}, \citenamefont {Gajewski}, \citenamefont {Gajos}, \citenamefont {Grudzie{\'n}}, \citenamefont {Hiesmayr} \emph {et~al.}}]{moskal2021positronium}%
  \BibitemOpen
  \bibfield  {author} {\bibinfo {author} {\bibfnamefont {P.}~\bibnamefont {Moskal}}, \bibinfo {author} {\bibfnamefont {K.}~\bibnamefont {Dulski}}, \bibinfo {author} {\bibfnamefont {N.}~\bibnamefont {Chug}}, \bibinfo {author} {\bibfnamefont {C.}~\bibnamefont {Curceanu}}, \bibinfo {author} {\bibfnamefont {E.}~\bibnamefont {Czerwi{\'n}ski}}, \bibinfo {author} {\bibfnamefont {M.}~\bibnamefont {Dadgar}}, \bibinfo {author} {\bibfnamefont {J.}~\bibnamefont {Gajewski}}, \bibinfo {author} {\bibfnamefont {A.}~\bibnamefont {Gajos}}, \bibinfo {author} {\bibfnamefont {G.}~\bibnamefont {Grudzie{\'n}}}, \bibinfo {author} {\bibfnamefont {B.~C.}\ \bibnamefont {Hiesmayr}}, \emph {et~al.},\ }\bibfield  {title} {\bibinfo {title} {Positronium imaging with the novel multiphoton pet scanner},\ }\href@noop {} {\bibfield  {journal} {\bibinfo  {journal} {Science Advances}\ }\textbf {\bibinfo {volume} {7}},\ \bibinfo {pages} {eabh4394} (\bibinfo {year} {2021})}\BibitemShut {NoStop}%
\bibitem [{\citenamefont {Shibuya}\ \emph {et~al.}(2020)\citenamefont {Shibuya}, \citenamefont {Saito}, \citenamefont {Nishikido}, \citenamefont {Takahashi},\ and\ \citenamefont {Yamaya}}]{shibuya2020oxygen}%
  \BibitemOpen
  \bibfield  {author} {\bibinfo {author} {\bibfnamefont {K.}~\bibnamefont {Shibuya}}, \bibinfo {author} {\bibfnamefont {H.}~\bibnamefont {Saito}}, \bibinfo {author} {\bibfnamefont {F.}~\bibnamefont {Nishikido}}, \bibinfo {author} {\bibfnamefont {M.}~\bibnamefont {Takahashi}},\ and\ \bibinfo {author} {\bibfnamefont {T.}~\bibnamefont {Yamaya}},\ }\bibfield  {title} {\bibinfo {title} {Oxygen sensing ability of positronium atom for tumor hypoxia imaging},\ }\href@noop {} {\bibfield  {journal} {\bibinfo  {journal} {Communications Physics}\ }\textbf {\bibinfo {volume} {3}},\ \bibinfo {pages} {173} (\bibinfo {year} {2020})}\BibitemShut {NoStop}%
\bibitem [{\citenamefont {Moskal}\ and\ \citenamefont {St{\k{e}}pie{\'n}}(2021)}]{moskal2021positronium_review}%
  \BibitemOpen
  \bibfield  {author} {\bibinfo {author} {\bibfnamefont {P.}~\bibnamefont {Moskal}}\ and\ \bibinfo {author} {\bibfnamefont {E.~{\L}.}\ \bibnamefont {St{\k{e}}pie{\'n}}},\ }\bibfield  {title} {\bibinfo {title} {Positronium as a biomarker of hypoxia},\ }\href@noop {} {\bibfield  {journal} {\bibinfo  {journal} {Bio-Algorithms and Med-Systems}\ }\textbf {\bibinfo {volume} {17}},\ \bibinfo {pages} {311} (\bibinfo {year} {2021})}\BibitemShut {NoStop}%
\bibitem [{\citenamefont {Shibuya}\ \emph {et~al.}(2022)\citenamefont {Shibuya}, \citenamefont {Saito}, \citenamefont {Tashima},\ and\ \citenamefont {Yamaya}}]{shibuya2022using}%
  \BibitemOpen
  \bibfield  {author} {\bibinfo {author} {\bibfnamefont {K.}~\bibnamefont {Shibuya}}, \bibinfo {author} {\bibfnamefont {H.}~\bibnamefont {Saito}}, \bibinfo {author} {\bibfnamefont {H.}~\bibnamefont {Tashima}},\ and\ \bibinfo {author} {\bibfnamefont {T.}~\bibnamefont {Yamaya}},\ }\bibfield  {title} {\bibinfo {title} {Using inverse laplace transform in positronium lifetime imaging},\ }\href@noop {} {\bibfield  {journal} {\bibinfo  {journal} {Physics in Medicine \& Biology}\ }\textbf {\bibinfo {volume} {67}},\ \bibinfo {pages} {025009} (\bibinfo {year} {2022})}\BibitemShut {NoStop}%
\bibitem [{\citenamefont {Conti}(2011)}]{conti2011improving}%
  \BibitemOpen
  \bibfield  {author} {\bibinfo {author} {\bibfnamefont {M.}~\bibnamefont {Conti}},\ }\bibfield  {title} {\bibinfo {title} {Improving time resolution in time-of-flight pet},\ }\href@noop {} {\bibfield  {journal} {\bibinfo  {journal} {Nuclear Instruments and Methods in Physics Research Section A: Accelerators, Spectrometers, Detectors and Associated Equipment}\ }\textbf {\bibinfo {volume} {648}},\ \bibinfo {pages} {S194} (\bibinfo {year} {2011})}\BibitemShut {NoStop}%
\bibitem [{\citenamefont {Cassidy}(2018)}]{cassidy2018experimental}%
  \BibitemOpen
  \bibfield  {author} {\bibinfo {author} {\bibfnamefont {D.~B.}\ \bibnamefont {Cassidy}},\ }\bibfield  {title} {\bibinfo {title} {Experimental progress in positronium laser physics},\ }\href@noop {} {\bibfield  {journal} {\bibinfo  {journal} {The European Physical Journal D}\ }\textbf {\bibinfo {volume} {72}},\ \bibinfo {pages} {1} (\bibinfo {year} {2018})}\BibitemShut {NoStop}%
\bibitem [{\citenamefont {Moskal}\ \emph {et~al.}(2024)\citenamefont {Moskal}, \citenamefont {Baran}, \citenamefont {Bass}, \citenamefont {Choinski}, \citenamefont {Chug}, \citenamefont {Curceanu}, \citenamefont {Czerwinski}, \citenamefont {Dadgar}, \citenamefont {Das}, \citenamefont {Dulski} \emph {et~al.}}]{moskal2024first}%
  \BibitemOpen
  \bibfield  {author} {\bibinfo {author} {\bibfnamefont {P.}~\bibnamefont {Moskal}}, \bibinfo {author} {\bibfnamefont {J.}~\bibnamefont {Baran}}, \bibinfo {author} {\bibfnamefont {S.}~\bibnamefont {Bass}}, \bibinfo {author} {\bibfnamefont {J.}~\bibnamefont {Choinski}}, \bibinfo {author} {\bibfnamefont {N.}~\bibnamefont {Chug}}, \bibinfo {author} {\bibfnamefont {C.}~\bibnamefont {Curceanu}}, \bibinfo {author} {\bibfnamefont {E.}~\bibnamefont {Czerwinski}}, \bibinfo {author} {\bibfnamefont {M.}~\bibnamefont {Dadgar}}, \bibinfo {author} {\bibfnamefont {M.}~\bibnamefont {Das}}, \bibinfo {author} {\bibfnamefont {K.~{\L}.}\ \bibnamefont {Dulski}}, \emph {et~al.},\ }\bibfield  {title} {\bibinfo {title} {First positronium image of the human brain in vivo},\ }\href@noop {} {\bibfield  {journal} {\bibinfo  {journal} {medRxiv}\ ,\ \bibinfo {pages} {2024}} (\bibinfo {year} {2024})}\BibitemShut {NoStop}%
\bibitem [{\citenamefont {Stepanov}\ \emph {et~al.}(2020)\citenamefont {Stepanov}, \citenamefont {Bokov}, \citenamefont {Ilyukhina},\ and\ \citenamefont {Byakov}}]{stepanov2020dissolved}%
  \BibitemOpen
  \bibfield  {author} {\bibinfo {author} {\bibfnamefont {S.~V.}\ \bibnamefont {Stepanov}}, \bibinfo {author} {\bibfnamefont {A.~V.}\ \bibnamefont {Bokov}}, \bibinfo {author} {\bibfnamefont {O.}~\bibnamefont {Ilyukhina}},\ and\ \bibinfo {author} {\bibfnamefont {V.~M.}\ \bibnamefont {Byakov}},\ }\bibfield  {title} {\bibinfo {title} {Dissolved oxygen and positronium atom in liquid media},\ }\href@noop {} {\bibfield  {journal} {\bibinfo  {journal} {Radioelektron. Nanosyst. Inf. Tehnol}\ }\textbf {\bibinfo {volume} {12}},\ \bibinfo {pages} {107} (\bibinfo {year} {2020})}\BibitemShut {NoStop}%
\bibitem [{\citenamefont {Hanggi}\ and\ \citenamefont {Carr}(1985)}]{hanggi1985errors}%
  \BibitemOpen
  \bibfield  {author} {\bibinfo {author} {\bibfnamefont {D.}~\bibnamefont {Hanggi}}\ and\ \bibinfo {author} {\bibfnamefont {P.~W.}\ \bibnamefont {Carr}},\ }\bibfield  {title} {\bibinfo {title} {Errors in exponentially modified gaussian equations in the literature},\ }\href@noop {} {\bibfield  {journal} {\bibinfo  {journal} {Analytical chemistry}\ }\textbf {\bibinfo {volume} {57}},\ \bibinfo {pages} {2394} (\bibinfo {year} {1985})}\BibitemShut {NoStop}%
\end{thebibliography}%
\end{document}